\documentclass[a4paper,10pt,twoside]{cpc-hepnp}

\usepackage{multicol}
\usepackage{graphicx}
\usepackage{booktabs}
\usepackage{amssymb,bm,mathrsfs,bbm,amscd}
\usepackage[tbtags]{amsmath}
\usepackage{lastpage}
\usepackage{CJK}

\begin{document}
\pdfoutput=1
\fancyhead[c]{\small submitted to 'Chinese Physics C' }
\fancyfoot[C]{\small 010201-\thepage}

\title{Study of helium irradiation induced hardening in MNHS steel\thanks{Supported by the National Basic Research Program of China (2010CB832902, 91026002) and National Natural Science
Foundation of China (U1232121) }}

\author{%
      Ji WANG$^{1,2,3;1)}$\email{wangji10@impcas.ac.cn}%
\quad Zhi-Guang WANG$^{1}$%
\quad Er-Qing Xie$^{1}$
\quad Ning GAO$^{1}$
\quad Ming-Huan CUI$^{1}$
\quad Kong-Fang WEI$^{1}$
\and Cun-Feng YAO$^{1}$
\quad Tie-Long SHEN$^{1}$
\quad Jian-Rong SUN$^{1}$
\quad Ya-Bin ZHU$^{1}$
\quad Li-Long PANG$^{1}$
\and Dong WANG$^{1,2}$
\quad Hui-Ping ZHU $^{1,2}$
\quad Yang-Yang DU$^{1,2}$
}
\maketitle

\address{%
$^1$ Institute of Modern Physics, Chinese Academy of Sciences, Lanzhou, 730000, China\\
$^2$ University of Chinese Academy of Sciences, Beijing, 100049, China\\
$^3$ School of Physical Science and Technology, Lanzhou University, Lanzhou, 730000, China\\
}

\begin{abstract}
A recently developed reduced activation ferritic/martensitic steel MNHS was irradiated with 200keV He ions to a fluence of ${\rm{1}} \times {\rm{1}}{{\rm{0}}^{{\rm{20}}}}{\rm{ions}}/{m^2}$ at 300$^\circ C$ and ${\rm{1}} \times {\rm{1}}{{\rm{0}}^{{\rm{21}}}}{\rm{ions}}/{m^2}$ at 300$^\circ C$ and 450$^\circ C$, respectively. The irradiation hardening of the steel was investigated by nanoindentation measurements combined with transmission electron microscopy (TEM) analysis. Dispersed barrier-hardening (DBH) model was applied to predict the hardness increments based on TEM analysis. The predicted hardness increments are consistent with the values obtained by nanoindentation tests. It is found that dislocation loops and He bubbles are hard barriers against dislocation motion and they are the main contributions to He irradiation-induced hardening of MNHS steel. The obstacle strength of He bubbles is stronger than the obstacle strength of dislocation loops.
\end{abstract}



\footnotetext[0]{\hspace*{-3mm}\raisebox{0.3ex}{$\scriptstyle\copyright$}2013
Chinese Physical Society and the Institute of High Energy Physics
of the Chinese Academy of Sciences and the Institute
of Modern Physics of the Chinese Academy of Sciences and IOP Publishing Ltd}%

\begin{multicols}{2}

\section{Introduction}

Reduced activation ferritic/martensitic steels are considered as one of the promising candidate structural materials for future fusion reactors because of their excellent mechanical properties, good thermal properties, low residual radioactivity and high swelling resistance [1]. Under the intense radiation fields in reactors, displacement damage and impurities (He and H) produced by transmutation reactions (i.e. (n, $\alpha$) and (n, p) reactions) are inevitable. Defects formed by accumulation of displaced atoms and the impurities such as dislocation loops, bubbles and clusters are thought to be obstacles to dislocation movement. As a result, structural materials served in reactors exhibit a hardness increase [2-7]. However, due to the nature of defect, the influences of different types of defects on the materials hardness are different. The role of dislocation loops and He bubbles among other types of irradiation-induced defects in materials hardness is still unclear. In order to investigate the effects of them on structural materials hardening, theoretic analysis based on dispersed barrier-hardening (DBH) model combined with TEM analysis and nanoindentation measurements were carried out to investigate the hardness changes of He implanted steels.

\section{Experimental}

The material used in this study is a recently developed reduced-activation ferritic/martensitic steel modified novel high silicon steel (MNHS). Chemical composition of the steel is listed in Table 1.
\begin{center}
\tabcaption{ \label{tab1}  Chemical composition (wt.$\% $) of the steel investigated in this study.}
\footnotesize
\begin{tabular*}{80mm}{c@{\extracolsep{\fill}}cccccccc}
\toprule Fe\hphantom{00}  & C & Cr & W & Mn & Si & V & Nb\\
\hline
Bal.\hphantom{00} & 0.25 & 10.78 & 1.19 & 0.54 & 1.42 & 0.19 & 0.01 \\
\bottomrule
\end{tabular*}
\vspace{0mm}
\end{center}
The steel was normalized at 1050$^\circ C$ for 30 minutes and tempered at 760$^\circ C$ for 90 minutes. Slices of $15 \times 15 \times {\rm{1m}}{{\rm{m}}^{\rm{3}}}$ were cut from the ingots after heat treatment. All specimens were prepared by mechanically polishing with SiC paper (up to 4000 grit) before polishing with diamond suspension (particle size $\sim $1$\mu $m). After mechanical polishing, all the specimens were electropolished to remove the work hardened surface.
Irradiation experiments were carried out in a terminal chamber of the 320kV multi-discipline research platform for highly charged ion at Institute of Modern Physics (IMP) in Lanzhou, China. The specimens were irradiated with 200keV He ions to a fluence of $1 \times {10^{20}}ions/{m^2}$ at 300$ \pm $5$^\circ C$ and $1 \times {10^{21}}ions/{m^2}$ at 300$ \pm $5$^\circ C$ and 450$ \pm $5$^\circ C$, respectively. Vacuum within the terminal chamber was maintained below ${\rm{5}} \times {\rm{1}}{{\rm{0}}^{{\rm{ - 5}}}}$Pa. Displacement damage and He concentration as a function of depth were calculated with SRIM code as shown in Fig.1.

\begin{center}
\includegraphics[width=8cm]{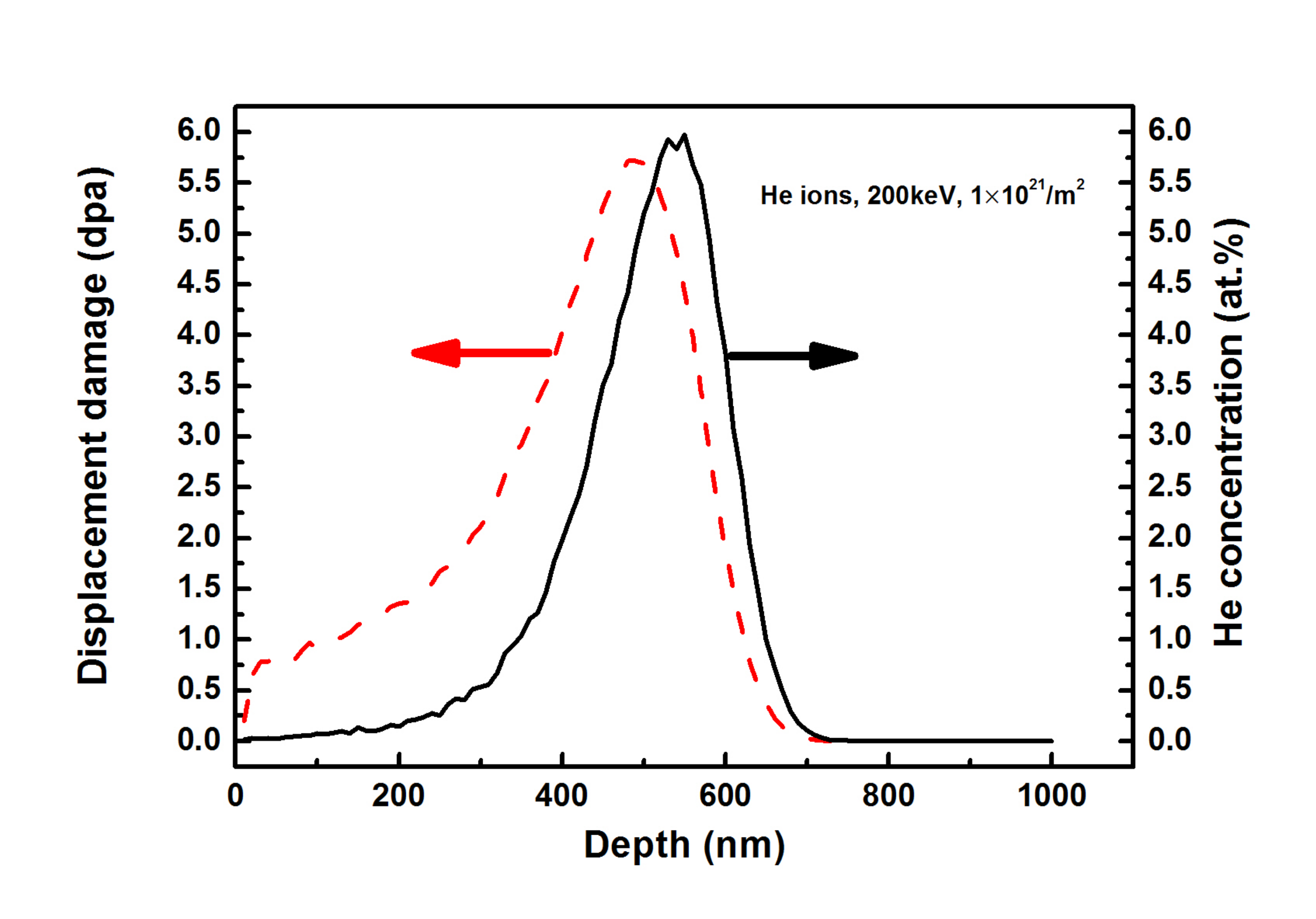}
\figcaption{\label{fig1}   Depth profile of displacement damage and He concentration. }
\end{center}

The displacement energy (Ed) was set to 40eV in SRIM calculation as recommended in ASTM E521-89 [8].
Nano-indentation tests were carried out using a Agilent Nano Indenter G200 with a Berkovich tip in a continuous stiffness mode (CSM). The indenter was normal to the samples surface. Each indentation was set approximately 30$\mu $m apart in order to avoid any overlap of the deformation region caused by other indentations. Care was taken to avoid uneven areas and areas where scratch, pitting and purities could be seen when labeling coordinates for indentations.
Cross-sectional TEM samples were prepared by ion milling in a Gatan precision ion polishing system after the indentation tests on the specimens finished. TEM analysis was performed in a FEI-TF20 transmission electron microscope operating at 200kV.

\section{Results and discussion}

\subsection{Characterization of He bubbles and dislocation loops in MNHS steel }

Microstructure of MNHS without He implantation is shown in Fig. 2.
 \begin{center}
\includegraphics[width=7cm]{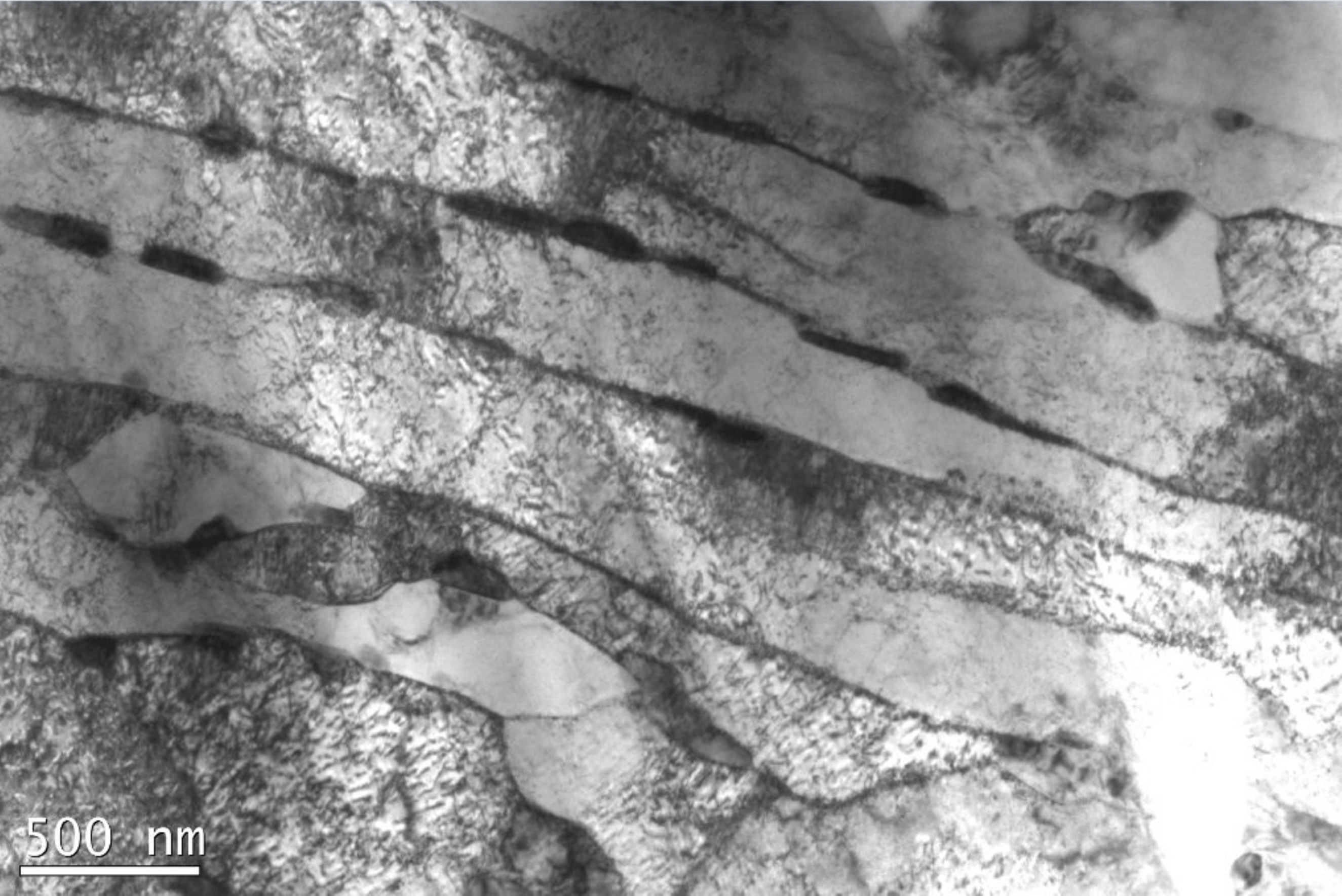}
\figcaption{\label{fig2}   TEM image of MNHS steel without He irradiation }
\end{center}
It can be seen that martensitic lath structures are the main features of this type of steel. Precipitations are found along lath boundaries. The average width of the martensitic lath is $ \sim $ 300nm. No dislocation loops could be detected.
After He ions irradiation, large numbers of bubbles and dislocation loops are found. Most of them distribute in a band region between the depth of $ \sim $ 400nm and $ \sim $ 600nm from surface. This is consistent with SRIM prediction as shown in Fig.1. Fig. 3 shows He bubbles in samples under different irradiation conditions.
\begin{center}
\includegraphics[width=4cm]{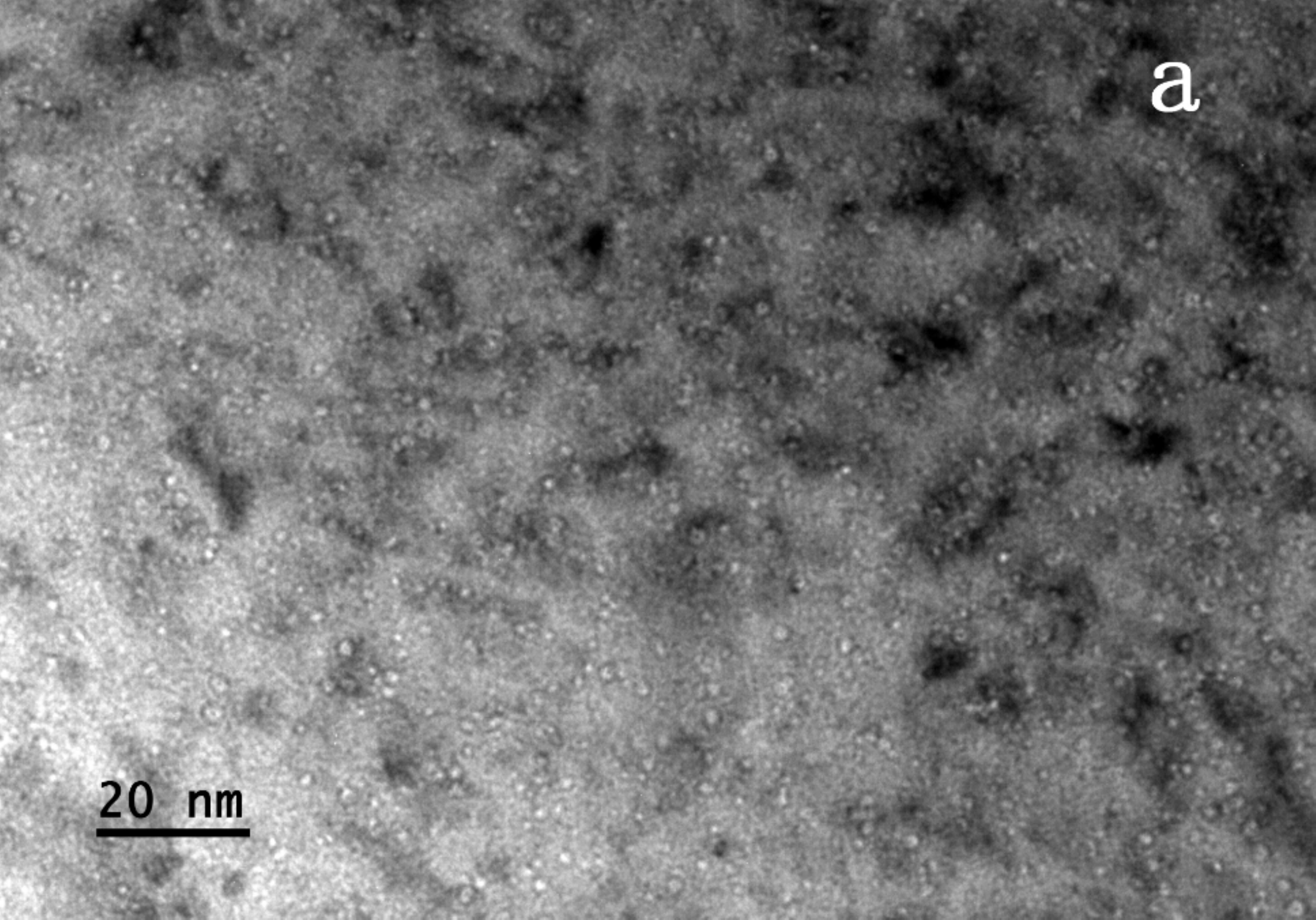}
\includegraphics[width=4cm]{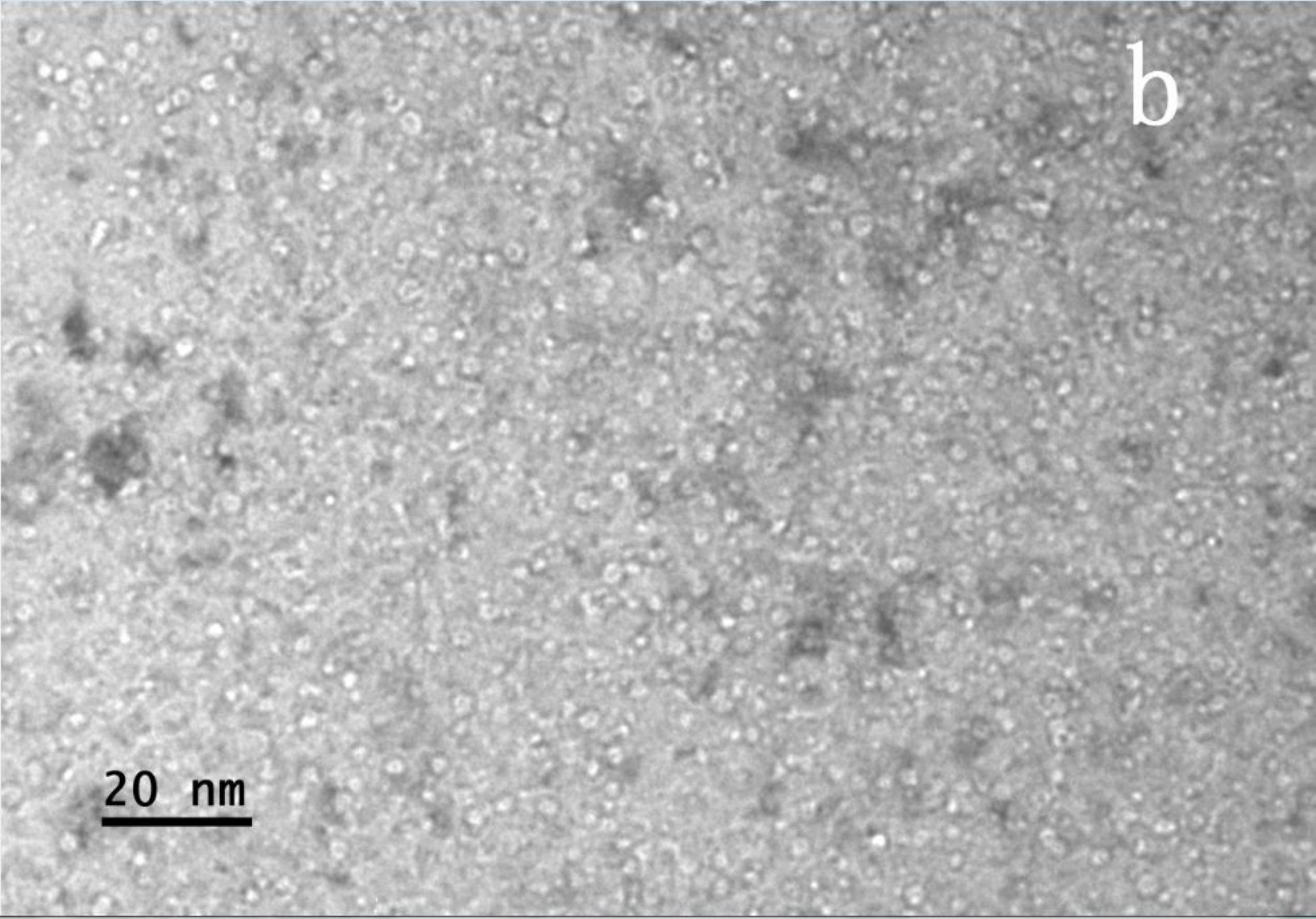}
\includegraphics[width=4cm]{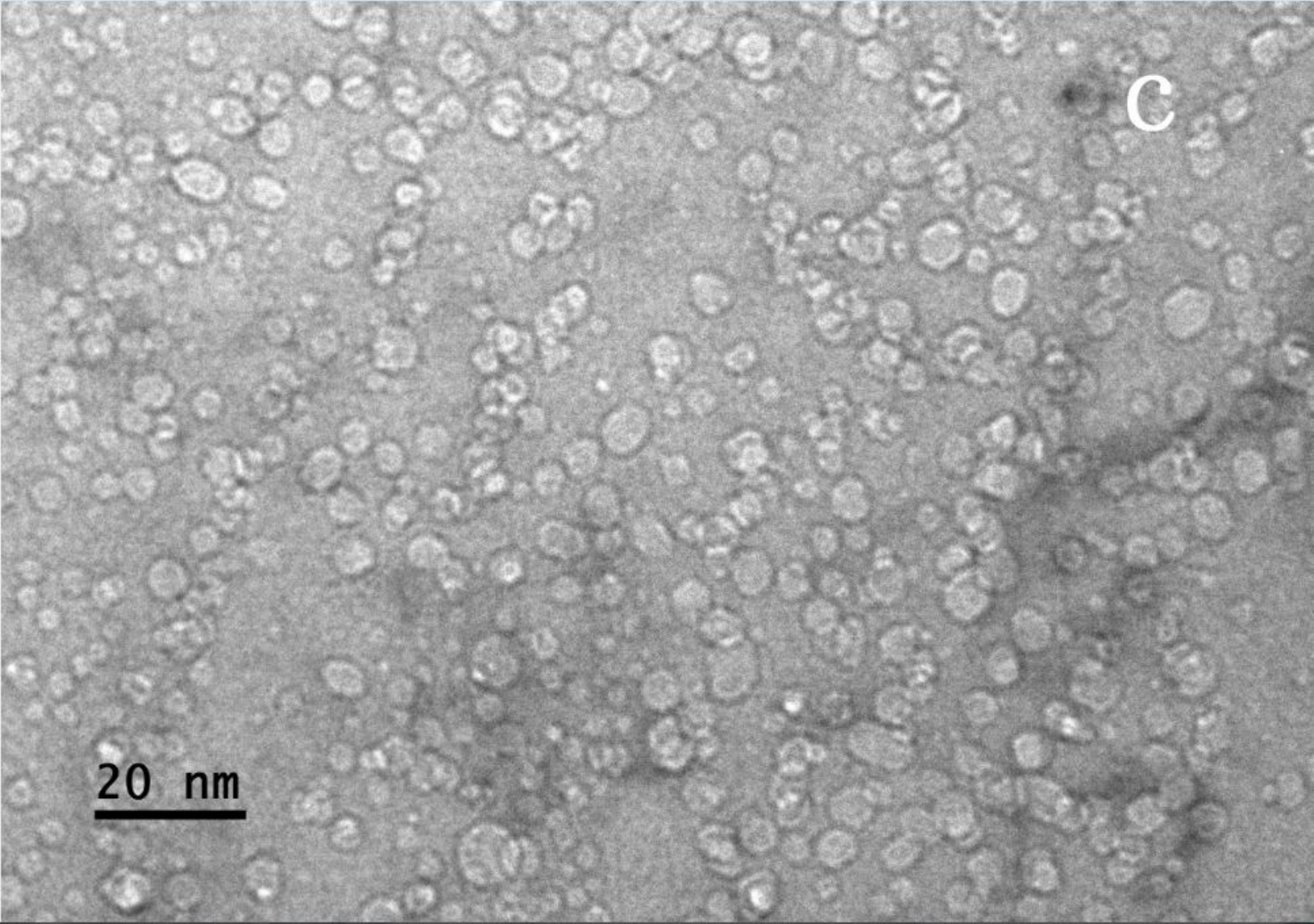}
\figcaption{\label{fig3}   TEM images of He bubbles in MNHS steels after He irradiation to (a)${\rm{1}} \times {\rm{1}}{{\rm{0}}^{{\rm{20}}}}{\rm{ions}}/{m^2}$ at 300$^\circ C$, (b)${\rm{1}} \times {\rm{1}}{{\rm{0}}^{{\rm{21}}}}{\rm{ions}}/{m^2}$ at 300$^\circ C$, (c)  ${\rm{1}} \times {\rm{1}}{{\rm{0}}^{{\rm{21}}}}{\rm{ions}}/{m^2}$ at 450$^\circ C$. }
\end{center} All the images were taken in the regions of the highest He concentration. Statistics on He bubbles size and number density were made and presented in Table 2.

\end{multicols}
\begin{center}
\tabcaption{ \label{tab2}  Number density(${\rho _b}$ for bubbles,${\rho _l}$ for dislocation loops), mean diameter($\overline {{R_b}} $ for bubbles,$\overline {{R_l}} $ for dislocation loops ) of dislocation loops and He bubbles in He irradiated MNHS steel.}
\footnotesize
\begin{tabular*}{170mm}{c@{\extracolsep{\fill}}cccccc}
\toprule Temperature($^\circ C$) & Dose (${\rm{ions}}/{m^2}$)   &$\overline {{R_l}} $ (nm)&$\overline {{R_b}} $(nm) &${\rho _l}$($/{m^3}$) &${\rho _b}$($/{m^3}$)\\
\hline
300 & ${\rm{1}} \times {\rm{1}}{{\rm{0}}^{{\rm{20}}}}{\rm{ions}}/{m^2}$ & 5.22 & 1.74&$2.4 \times {10^{22}}$ & $2.17 \times {10^{23}}$ \\
300 & ${\rm{1}} \times {\rm{1}}{{\rm{0}}^{{\rm{21}}}}{\rm{ions}}/{m^2}$ & 7.06 & 2.34&$9.85 \times {10^{22}}$ & $4.73 \times {10^{23}}$ \\
450 & ${\rm{1}} \times {\rm{1}}{{\rm{0}}^{{\rm{21}}}}{\rm{ions}}/{m^2}$ & 11.45 & 3.56& $ 8.76 \times {10^{21}} $ & $5.91 \times {10^{22}}$ \\
\bottomrule
\end{tabular*}%
\end{center}
\begin{multicols}{2}It can be seen that the number density of bubbles increases as the increase of irradiation dose and decreases as the irradiation temperature increases while the size of bubbles increases when irradiation dose and irradiation temperature increase, respectively. Large amount of dislocation loops are also found in high helium concentration region as shown in Fig. 4. Statistics made on the size and number density of dislocation loops are also shown in Table 2, which indicates that the size of dislocation loops increases as the increasing of irradiation temperature and dose, respectively.

\begin{center}
\includegraphics[width=4cm]{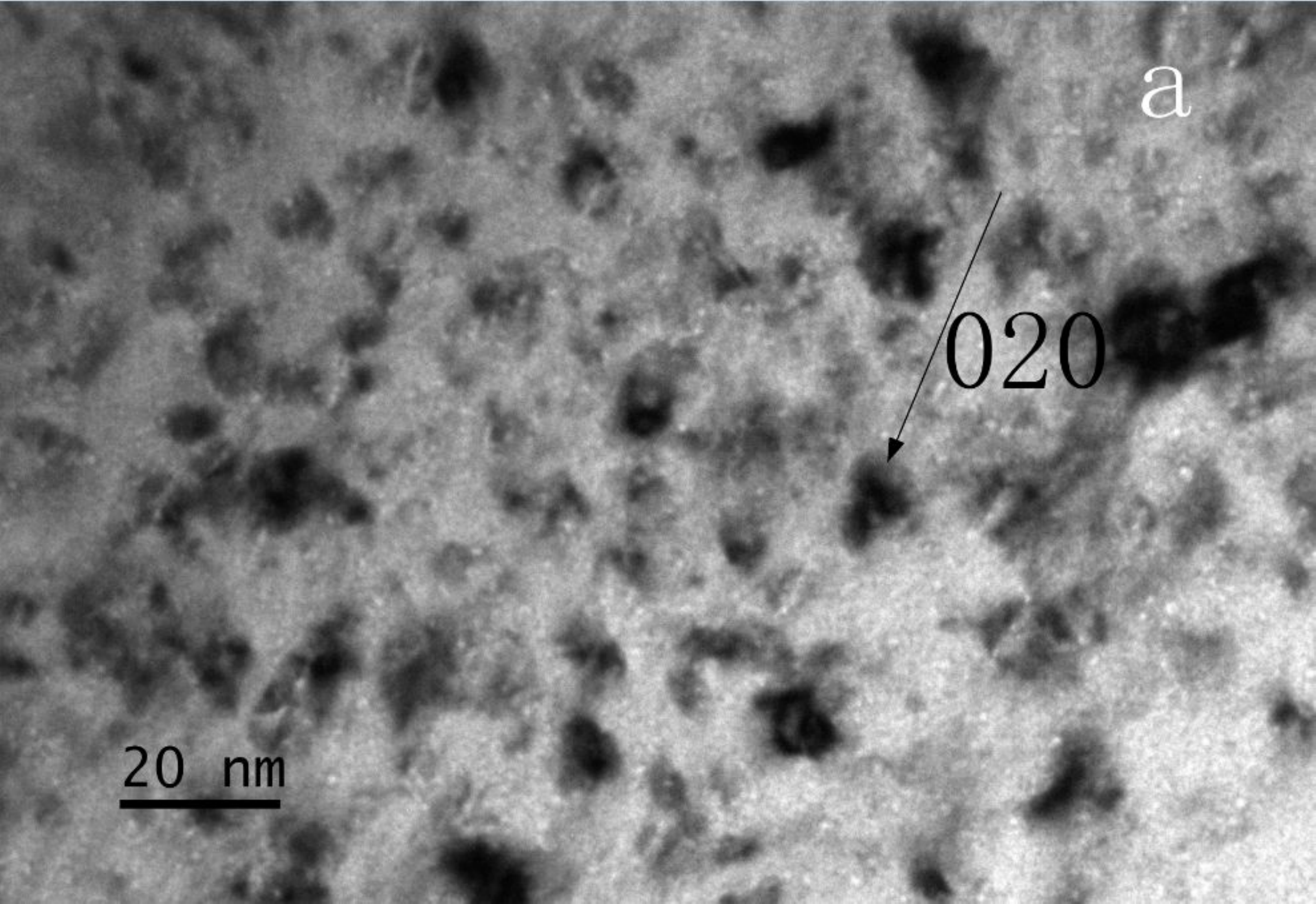}
\includegraphics[width=4cm]{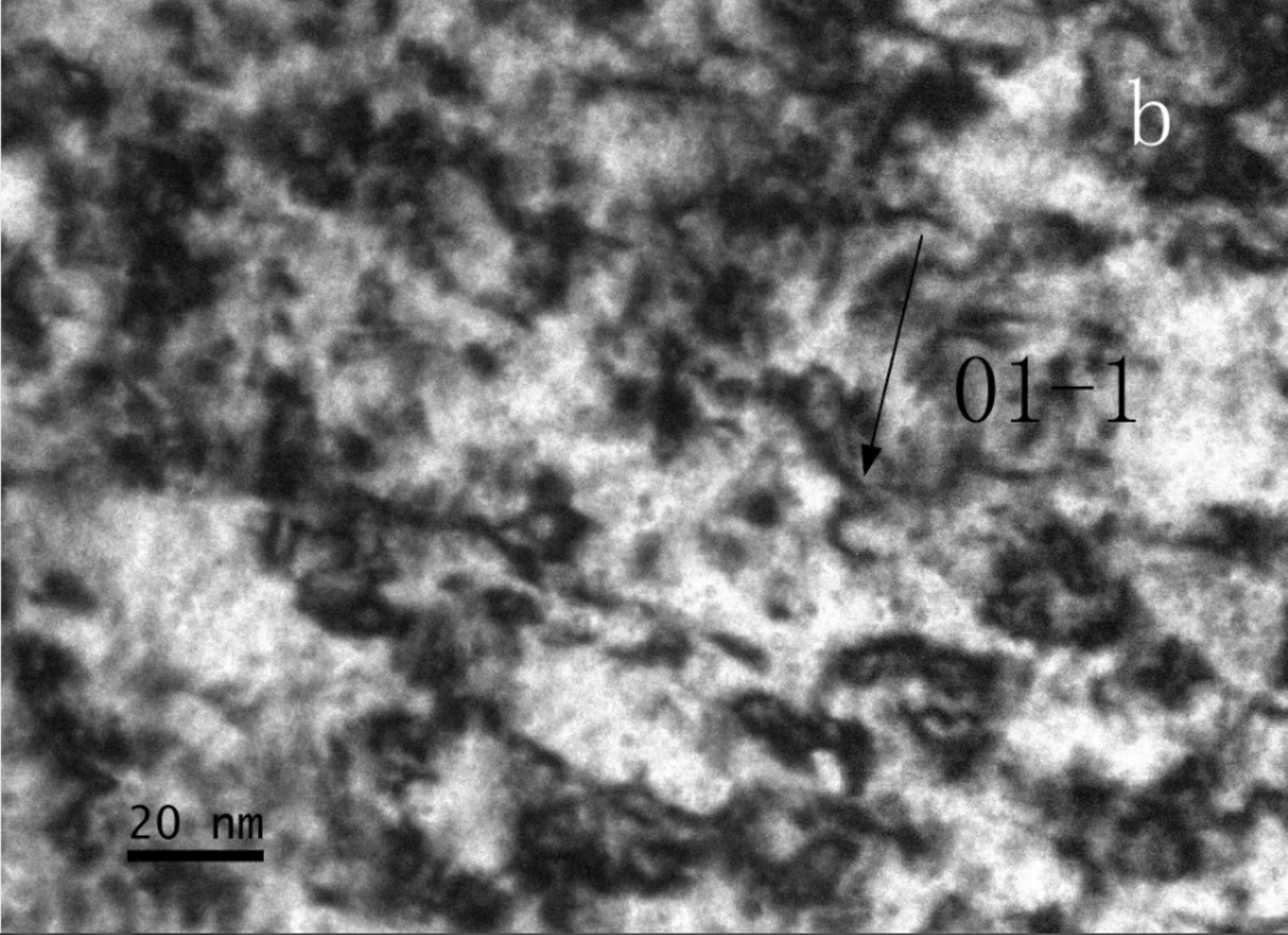}
\includegraphics[width=4cm]{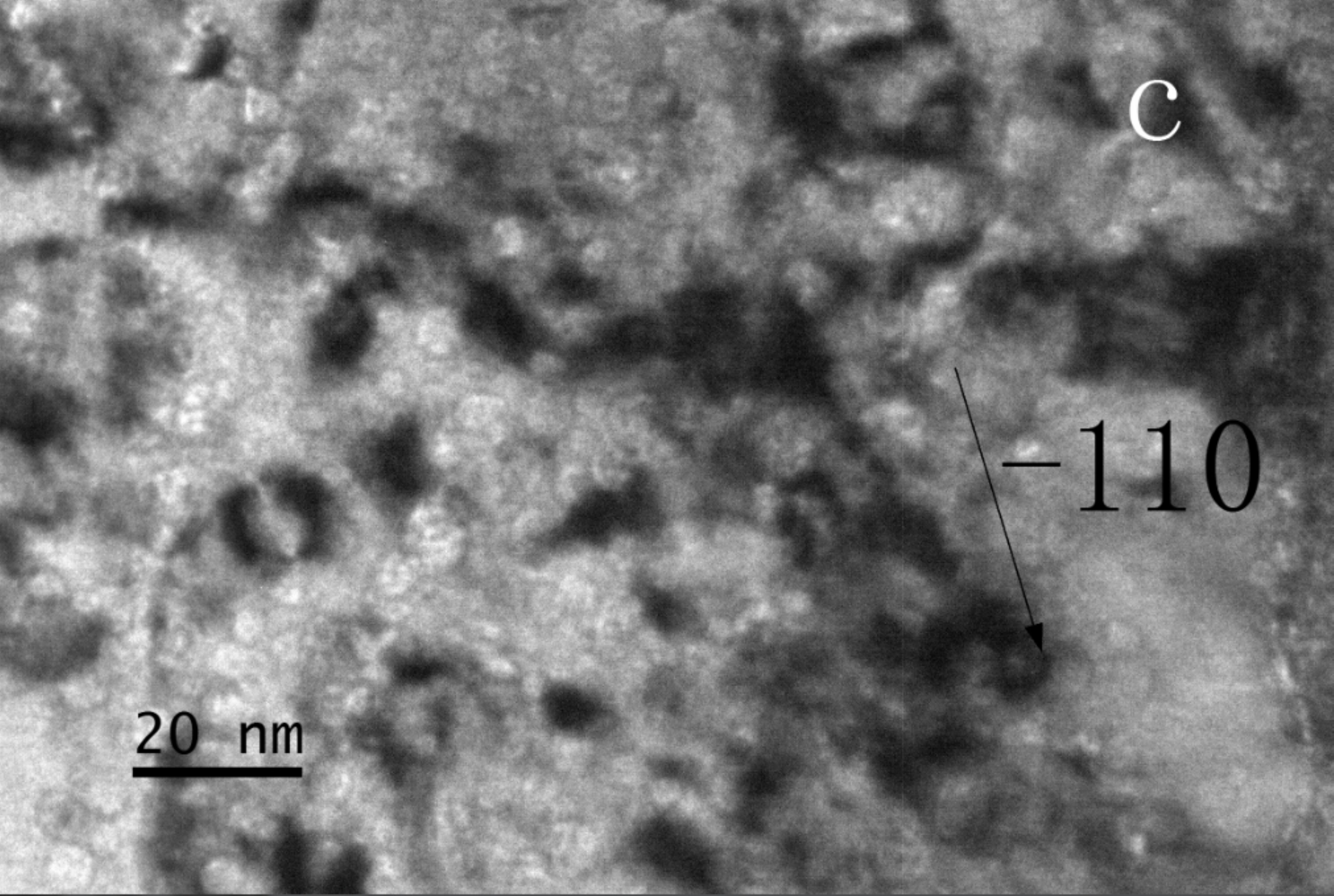}
\figcaption{\label{fig4}   TEM images of dislocation loops in MNHS steels after He irradiation to  (a)${\rm{1}} \times {\rm{1}}{{\rm{0}}^{{\rm{20}}}}{\rm{ions}}/{m^2}$ at 300$^\circ C$, (b)${\rm{1}} \times {\rm{1}}{{\rm{0}}^{{\rm{21}}}}{\rm{ions}}/{m^2}$ at 300$^\circ C$, (c)  ${\rm{1}} \times {\rm{1}}{{\rm{0}}^{{\rm{21}}}}{\rm{ions}}/{m^2}$ at 450$^\circ C$. }
\end{center}

\subsection{Nano-indentation tests}

The values of hardness in this study were obtained by averaging six individual indentation results on each sample. Occasionally, a hardness versus depth curve would be far away from the average values, which may be due to a simple failure during the indentation test or the subsurface precipitates near the indenter tip. Such outliers were removed in this work. Due to the uncertainty of indenter geometry and testing artifacts, hardness data at the depth $ < $50nm from surface were not reliable. Thus, we took the values at the depth over 50 nm for analysis. Fig. 5 shows the average hardness as a function of indentation depth for irradiated samples and an unirradiated sample.
\begin{center}
\includegraphics[width=8cm]{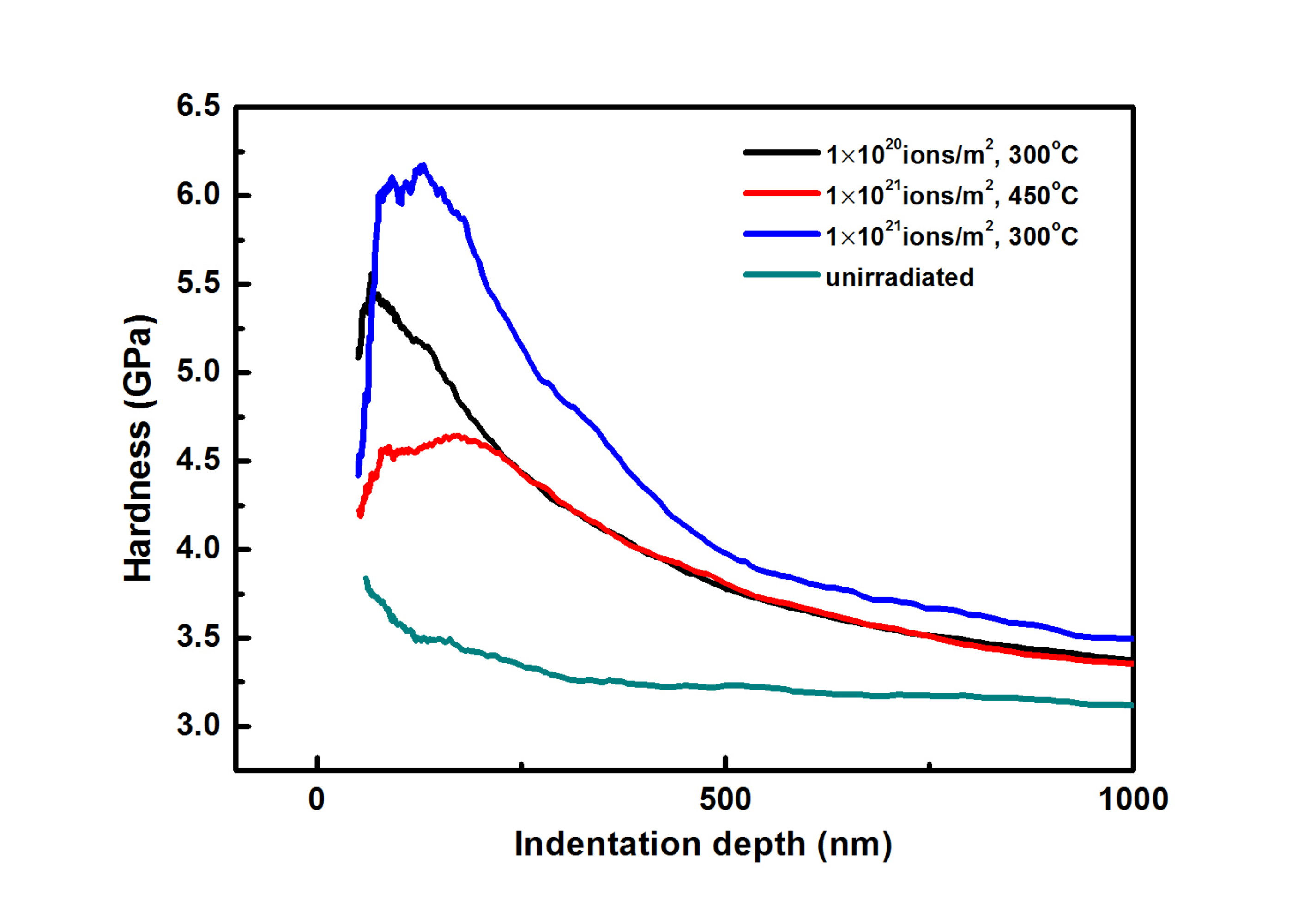}
\figcaption{\label{fig5}   Indentation depth profile of hardness of irradiated and unirradiated MNHS steels.  }
\end{center}
For the irradiated samples, a hardness peak is observed at depth of $ \sim $ 100 nm, different from the irradiation damage peak of $ \sim $ 500 nm. This is consistent with previous work [5,9,10] that the depth of damage peak is 5-7 times of the depth of the hardness peak. The reason for above difference is probably due to the fact that the depth of a plastic zone with an approximately hemispherical shape under the indentation reaches about 5-7 times of the depth at which  indenter tip can reach[7,10]. At the depth of hardness peak, the indenter was mostly sampling the He implanted region and the surface effects were reduced. Therefore, the hardness peak in fact indicates the effects of helium implantation region. We therefore take the hardness values at the depth of 100nm for comparing the relative hardening among samples. The relative hardness increments of He implanted samples obtained by nanoindentation tests ($\Delta {H_e}$) were summarized in table 3.
\begin{center}
\tabcaption{ \label{tab3}  Comparison of predicted hardness increment($\Delta {H_p}$) and experimentally obtained hardness increment($\Delta {H_e}$).}
\footnotesize
\begin{tabular*}{80mm}{c@{\extracolsep{\fill}}cccc}
\toprule Temperature(${}^oC$)  &Dose ($ions/{m^2}$) &$\Delta {H_p}$(GPa) & $\Delta {H_e}$(GPa) \\
\hline
300 &$1 \times {10^{20}}$ &1.50 &1.68 \\
300 &$1 \times {10^{21}}$ &2.71 &2.61 \\
450 &$1 \times {10^{21}}$&1.15 &1.18 \\
\bottomrule
\end{tabular*}
\vspace{0mm}
\end{center}

\subsection{Analysis of irradiation induced hardening}
Structural materials for fusion reactors exhibit radiation-induced hardening when irradiated with neutron or energetic ions [2-7, 11]. This hardening is thought to be caused by defects formed during irradiation such as clusters of interstitial atoms, small bubbles, precipitates and dislocation loops [11-14]. Previous studies showed that dislocation loops and bubbles are strong obstacles that act as barriers to dislocation motion, resulting in the increase of hardness of irradiated materials [3,4,6,14,15]. In order to have a better understanding of the hardness changes caused by dislocation loops and He bubbles. An attempt is made below to analyses the hardness increase based on dispersed barrier-hardening (DBH) model.
As the interstitial dislocation loops and He bubbles are classified as short-range obstacles [2].  The increase of yield strength due to one type of short-range obstacle could be expressed as

\begin{eqnarray}
\label{eq1}
\Delta {\sigma _y} = M\alpha \mu b{\left( {Nd} \right)^{{1 \mathord{\left/
 {\vphantom {1 2}} \right.
 \kern-\nulldelimiterspace} 2}}}.
\end{eqnarray}

Where$\Delta {\sigma _y}$ is the yield strength increment, M is the Taylor factor, $\alpha$ is the barrier strength factor, $\mu $ is the shear modulus, b is the Burgers vector, N and d are number density and mean size of obstacles, respectively. For two types of short-range strong obstacles, the superposition rule could be expressed as following
\begin{eqnarray}
\label{eq2}
\Delta {\sigma ^2} = \Delta \sigma _1^2 + \Delta \sigma _2^2.
\end{eqnarray}

\begin{eqnarray}
\label{eq3}
\Delta \sigma _1^2 = M{\alpha _1}\mu b{\left( {{N_1}{d_1}} \right)^{{\raise0.7ex\hbox{$1$} \!\mathord{\left/
 {\vphantom {1 2}}\right.\kern-\nulldelimiterspace}
\!\lower0.7ex\hbox{$2$}}}}.
 \end{eqnarray}

\begin{eqnarray}
\label{eq4}
\Delta \sigma _2^2 = M{\alpha _2}\mu b{\left( {{N_2}{d_2}} \right)^{{\raise0.7ex\hbox{$1$} \!\mathord{\left/
 {\vphantom {1 2}}\right.\kern-\nulldelimiterspace}
\!\lower0.7ex\hbox{$2$}}}}.
\end{eqnarray}

M is 3.06 for bcc metals, the shear modulus for MNHS steel is 80GPa and b=$2.48 \times {10^{ - 10}}$ m. The number density and mean size of dislocation loops and bubbles in three sets of samples under different He ions irradiation conditions are summarized in Table 2. According to previous work, dislocation loops and bubbles are thought as strong obstacles with the $\alpha $ of 0.25 - 0.5 and 0.3 - 0.5 [3,6,15], respectively (some work show higher values of $\alpha $). We apply above values (with $\alpha $ values of 0.25 for dislocation loops and 0.4 for He bubbles) in equations (2) to (4), then the predicted increments of yield strength ($\Delta {\sigma _y}$) were got for steels irradiated under 200keV He ions.
\\
\indent It is reported that the correlation between change in hardness and change in yield stress was determined to be [16].
\begin{eqnarray}
\label{eq5}
\Delta {\rm{H}} = 3\Delta {\sigma _y}.
\end{eqnarray}

Using this relationship, the predicted hardness increment ($\Delta {{\rm{H}}_{\rm{p}}}$) could be calculated based on DBH model. It is thus possible to compare the predicted hardness changes with hardness changes obtained by nanoindentation tests. The comparison between predicted hardness changes ($\Delta {{\rm{H}}_{\rm{p}}}$) and experimentally obtained hardness changes ($\Delta {{\rm{H}}_{\rm{e}}}$) is summarized in Table 3. It can be seen that the predicted changes of hardness based on DBH model are consistent with hardness changes obtained by nanoindentation tests. This suggests that both dislocation loops and He bubbles are the main contributions to irradiation-induced materials hardening and dislocation loops and He bubbles are strong pinning centers with the pinning strength of He bubbles ($\alpha $=0.4) stronger than the pinning strength of dislocation loops ($\alpha $=0.25). However, a note should be mentioned about the rates of introduction of displacement damage and helium. It is known that radiation-induced defects accumulate differently under different damage and helium production rates [17]. As a result, mechanical properties of irradiated structural materials vary accordingly [13]. In this experiment, the damage rate and introduction rate of helium are $6.39 \times {10^{ - 4}}$ dpa/s and 6.67appm/s, respectively.

\section{Conclusions}

The effects of helium under conditions of simultaneous displacement damage production on the hardness of ferritic/martensitic steel MNHS were studied by He implantation into MNHS steel combined with nano-indentation technique and TEM analysis. The DBH model was adopted to predict the hardness changes caused by dislocation loops and He bubbles. it is found that the predicted hardness changes based on DBH model are consistent with the experimentally obtained hardness changes. Dislocation loops and He bubbles are hard obstacles against dislocation motion and they are the main contributions to He irradiation-induced hardening of MNHS steel. The barrier strength of He bubbles ($\alpha $=0.4) is stronger than the barrier strength of dislocation loops ($\alpha $=0.25).

\acknowledgments{This work was supported by the National Basic Research Program of China (2010CB832902, 91026002), National Natural Science Foundation of China (U1232121). The authors greatly appreciate all the help from staff of 320kV multi-discipline research platform for highly charged ions during the implantation experiments.}

\end{multicols}

\vspace{15mm}

\begin{multicols}{2}

\end{multicols}

\clearpage

\end{document}